\documentclass [12pt]{article}
\usepackage{graphicx}
\begin{document}
\vskip 9cm
\begin{center}
\Large
{\bf Determination of Spacecraft Attitude and Source Position Using
Non-aligned Detectors in Spin-stabilized Satellites}\\
\end{center}
\normalsize
\begin{center}
{Srikanta Sinha}\\
{S.N. Bose National Centre for Basic Sciences}\\
{Sector III, Block JD, Salt Lake, Kolkata-700 098, INDIA}
\end{center}
\vskip 2cm
\section{Abstract}
The modulation of high-energy transients' (or steadily emitting sources') light curves
due to the imperfect alignment of the detector's view axis with the spin
axis in a spin-stabilized satellite is derived. It is shown
how the orientation of the detector's view axis with respect to the satellite's
spin axis may be estimated using observed light curves. The effects of
statistical fluctuations are considered.

Conversely, it is shown how the attitude of a spin-axis stabilized satellite
as well as the unknown position of a celestial source of high-energy photons
may be determined using a detector whose view-axis is intentionally kept
inclined and is known accurately beforehand. The case of three-axes 
stabilized satellites is also discussed.
\vskip 1cm
Keywords: space vehicles:instruments, methods:analytical
%\vskip 20cm
\vskip 2cm
\section{Introduction:}
In high-energy (X-ray and gamma ray) astronomy experiments photon 
detectors (scintillators such as Sodium Iodide, Cesium Iodide or gas-filled
multi-wire proportional counters (MWPCs) or even solid-state detectors
such as Si(PIN) or CdTe detectors) are flown on-board satellites. The
satellites are either spin-axis stabilized (such as the Indian SROSS C-2
satellite) or three-axes stabilized. In the case of spin-axis stabilized
satellites, normally, the detector's view axis is aligned with the 
spin-axis of the satellite. However, sometimes this alignment is not
perfect and the view axis of the detector makes a small but finite angle
with the spin-axis of the satellite. In this case the observed light
curve of a given celestial source of photons is modulated by the spin
-period of the satellite. The amplitude of the modulation is a function
of the polar angle ($\theta_{v}$) between the detector's view axis and the satellite's
spin axis. It also depends on the relative direction of the source with 
respect to the spin axis ($\theta_{0}^{'}$).

The spin-period is usually not strictly constant over the entire life
time of the satellite. But over a small time period (during the entire
duration of a high energy transient) it may be assumed to be almost
constant.

In the present paper we derive the amplitude of the
light curve's modulation as a function of the space angle $\theta_{v}$.
We take a high energy transient light curve and superimpose the
modulation (using a given value of $\theta_{v}$). 
We show how the (if unknown) orientation of the view axis may be estimated
reasonably accurately using the observed modulated light curve and the
unmodulated light curve of the same high energy transient (say, a Solar
X-ray flare) detected by another satellite. 

Finally, we discuss how this apparently disadvantageous and undesirable
phenomenon may be useful in determining the position of a celestial
high energy photon source given that the orientation of the detector's
view axis (intentionally kept inclined) is known beforehand. Also it is
described how this same phenomenon might be utilised in determining the
(unknown) attitude of the spacecraft by placing two small detectors (one
inclined to the spin-axis, the other aligned with the spin-axis).
%\section{Method of Calculations:}
\section{Mathematical Formulation:}
Let $XYZ$ (Fig.1) denote the celestial equatorial co-ordinate system. Let $\alpha_{S}$
and $\delta_{S}$ denote the orientation of the satellite's spin axis in the
$XYZ$ system. We define $\theta_{S} = \pi -\delta_{S}$ and $\phi_{S} = \alpha_{S}$.
Let the source direction 
be $\alpha_{0}$ and $\delta_{0}$ respectively.
\begin{figure}[htp]
\includegraphics
[height=5cm,width=6.5cm,angle=0]{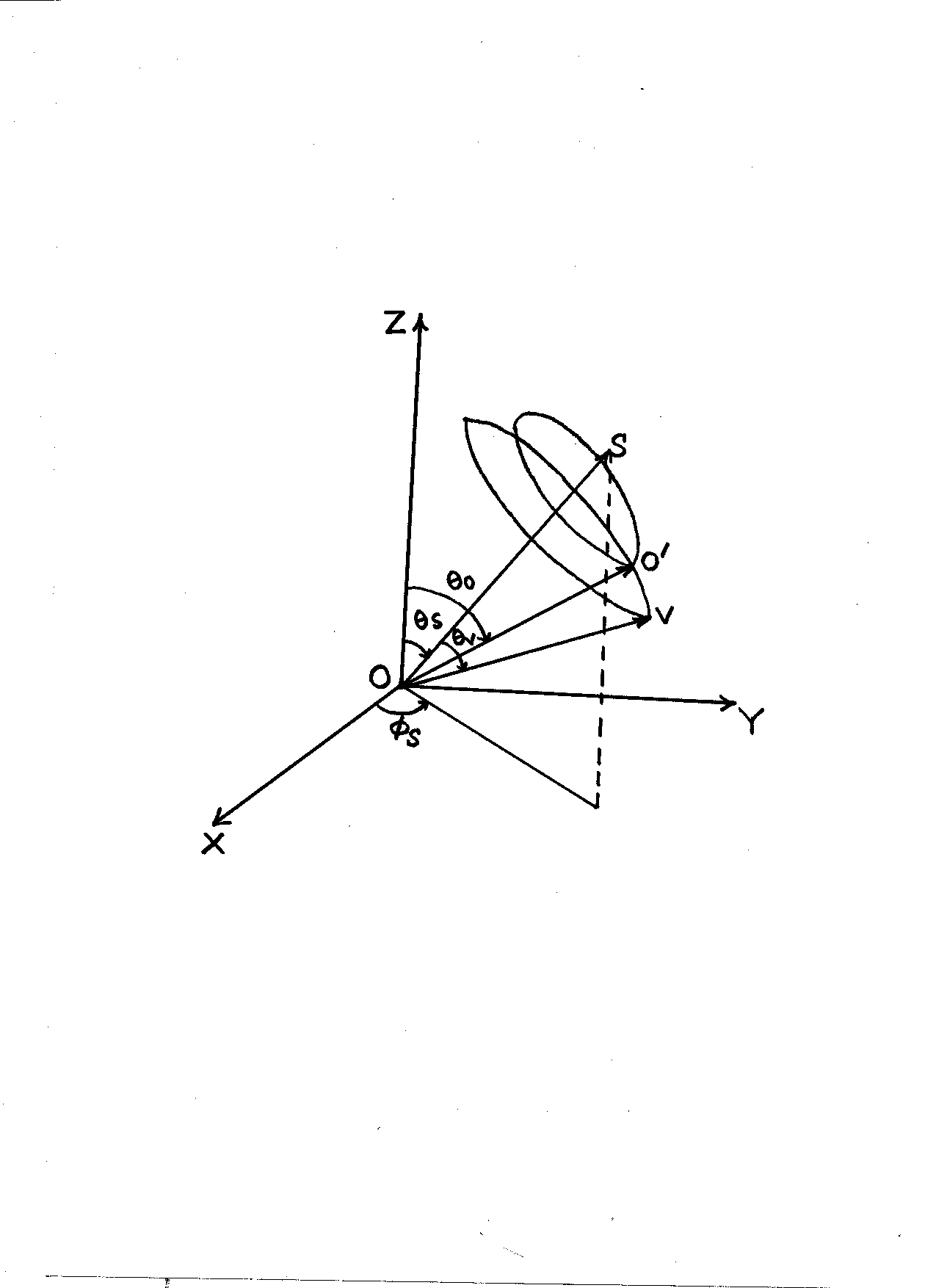}
\caption{XYZ is the celestial polar co-ordinate system. OS is the
spin-axis of the satellite. OO' is the vector in the source direction.
OV is the detector's view axis. The detector's view axis and the source
rotate around the spin axis at the spin period of the satellite.}
\end{figure}
\vskip 0.2cm
Then, $\theta_{0} = \pi - \delta_{0}$ and $\phi_{0} = \alpha_{0}$. The
direction cosines of the source 
with respect to the spin-axis system may be obtained as
\begin{equation}
\left(\matrix {sin\theta_{0}^{'}cos\phi_{0}^{'}\cr
sin\theta_{0}^{'}sin\phi_{0}^{'}\cr
cos\theta_{0}^{'}}\right) =
\left(\matrix {cos\theta_{S}cos\phi_{S} & cos\theta_{S}sin\phi_{S} & -sin\theta_{S}\cr
-sin\phi_{S} & cos\phi_{S} & 0\cr
sin\theta_{S}cos\phi_{S} & sin\theta_{S}sin\phi_{S} & cos\theta_{S}}\right)
\left(\matrix {sin\theta_{0}cos\phi_{0}\cr
sin\theta_{0}sin\phi_{0}\cr
cos\theta_{0}}\right)
\end{equation}
(see, for example, Arfken and Weber).
Hence the unit vector in the source direction is 
\begin{equation}
\vec S = \vec i^{'} sin\theta_{0}^{'} cos\phi_{0}^{'}
+\vec j^{'} sin\theta_{0}^{'} sin\phi_{0}^{'} +\vec k^{'} cos\theta_{0}^{'}
\end{equation}
Let the direction of the detector's view axis be $\theta_{v}$ and
$\phi_{v}$ respectively with respect to the spin-axis co-ordinate
system. 
Then the unit vector along the view axis in the spin-axis system
is given by
\begin{equation}
\vec V = \vec i^{'} sin\theta_{v}cos\phi_{v}+
\vec j^{'} sin\theta_{v} sin\phi_{v} +\vec k^{'} cos\theta_{v}
\end{equation}
If we assume $\phi_{v} =0$, this becomes
\begin{equation}
\vec V = \vec i^{'} sin\theta_{v} + \vec k^{'} cos\theta_{v}
\end{equation}
This essentially implies that the spin-axis system is given a
trivial rotation by an amount $\phi_{v}$ in the opposite sense.
\subsection{Determination of the inclination of the view axis:}
The modulation factor (assuming $\phi_{v}=0$) is given by
\begin{equation}
M=sin\theta_{v}sin\theta_{0}^{'}cos(\omega t+\psi)+
cos\theta_{v}cos\theta_{0}^{'}
\end{equation}
where we have put $\phi_{0}^{'}=(\omega t+\psi)$, $\psi$ being the
epoch, $t$ the time and $\omega$ the angular velocity due to the
spin.

The modulation affects only the signal and not the background counts.
The modulated counts
\begin{equation}
D_{i}=C_{i} M =C_{i} (sin\theta_{v}sin\theta_{0}^{'}
cos(\omega t_{i} +\psi)+cos\theta_{v}cos\theta_{0}^{'})
\end{equation}
where the $C_{i}$'s are the counts in the unmodulated 
light-curve.

To determine $\theta_{v}$ ($\theta_{0}^{'}$ is given) one has to
consider another unmodulated light curve (this light curve
has to be multiplied by an appropriate factor $g$ such that the
maxima (peaks) are the same for the two light curves.

Therefore, we have
\begin{equation}
sin\theta_{v}sin\theta_{0}^{'}cos(\omega t_{i}+\psi)+
cos\theta_{v}cos\theta_{0}^{'} = f_{i}
\end{equation}
where$f_{i}=D_{i}/C_{i}, C_{i}$ being known from a detector
which is aligned (on-board a different satellite). Here,
of course, the modulated light curve has to be multiplied
by a suitable factor such that the maxima of the two light
curves are equal.
This may be written as (since $\theta_{0}^{'}$ is known, i.e. both
$sin\theta_{0}^{'}$ and $cos\theta_{0}^{'}$ are known).
\begin{equation}
a_{1}x_{1}cos(\omega t_{i}+\psi) + b_{1} x_{2} = f_{i}
\end{equation}
where $x_{1}=sin\theta_{v}, x_{2}=cos\theta_{v}$.
Also, $a_{1}=sin\theta_{0}^{'}$, $b_{1}= cos\theta_{0}^{'}$.
Taking the time average of both sides (over an integral number
of cycles),
\begin{equation}
<a_{1}x_{1}cos(\omega t_{i}+\psi)>+<b_{1}x_{2}>=<f_{i}>
\end{equation}
\begin{equation}
a_{1}x_{1}<cos(\omega t_{i}+\psi)>+b_{1}x_{2}=<f_{i}>
\end{equation}

Since $<cos(\omega t_{i}+\psi)>$ is equal to zero for an integral
number of cycles, the first term becomes equal to zero.
Hence,
\begin{equation}
b_{1}x_{2}=<f_{i}>
\end{equation}
or,
\begin{equation}
x_{2}=(1/b_{1})<f_{i}>
\end{equation}
Since, $x_{2}=cos\theta_{v}$, $\theta_{v}=cos^{-1}(1/b_{i})<f_{i}>$.
\subsection{Determination of Source Location}
If $\theta_{v}$ is known beforehand, it is possible to determine the location
of a source in the sky using two detectors, one aligned, the other inclined
(at a known angle) with the spin axis of the satellite.

The equation for the modulated light curve (counts vs. time) is
\begin{equation}
D_{i}=C_{i}(sin\theta_{v}cos\phi_{v}sin\theta_{0}^{'}cos\phi_{0}^{'}
+sin\theta_{v}sin\phi_{v}sin\theta_{0}^{'}sin\phi_{0}^{'}
+cos\theta_{0}^{'}cos\theta_{v})
\end{equation}
where $C_{i}$ is the counts detected during the ith time bin (when the
view axis is aligned with the spin axis, i.e. $\theta_{v}=0$).
The other symbols have their usual
meanings.
Assuming $\phi_{v}=0$ (reorientation of the spin-axis system), this
reduces to
\begin{equation}
D_{i}=C_{i}(sin\theta_{v}sin\theta_{0}^{'}cos\phi_{0}^{'}
+cos\theta_{0}^{'}cos\theta_{v})
\end{equation}
Here, $\phi_{0}^{'}=cos(\omega t_{i}+\psi)$, $\psi$ being the epoch.
For a detector whose view axis is aligned with the spin axis,
$\theta_{v}=0$. Then
\begin{equation}
D_{i}=C_{i}(cos\theta_{0}^{'})
\end{equation}
From the last equation, for a given (say, the ith) time bin,
\begin{equation}
cos\theta_{0}^{'}=D_{i}/C_{i} 
\end{equation}
which gives
\begin{equation}
\theta_{0i}^{'}=cos^{-1}(D_{i}/(C_{i}))
\end{equation}
Let $a=cos\theta_{0}^{'}$ (known) and $sin\theta_{0}^{'}=\pm \sqrt(1-a^{2})$.
If, for a second detector (for which $\theta_{v} \ne 0$, i.e. $sin\theta_{v} \ne 0$,
then
\begin{equation}
D_{i}^{'}=C_{i}^{'}(br\phi_{0}^{'}+ac)
\end{equation}
This gives
\begin{equation}
\phi_{0i}^{'}=cos^{-1}((1/br)(D_{i}^{'}/C_{i}^{'}-ac))
\end{equation}
Here $b=sin\theta_{v}$, $r=sin\theta_{v}^{'}$ and $c=cos\theta_{v}$.
Knowing the values of $\theta_{0}^{'}$ and $\phi_{0}^{'}$, the inverse
transformation (corresponding to that given in eqn. (1)) may be used in order to
obtain the values of $\theta_{0}$ and $\phi_{0}$.
\begin{equation}
\left(\matrix {sin\theta_{0}cos\phi_{0}\cr
sin\theta_{0}sin\phi_{0}\cr
cos\theta_{0}}\right) =
\left(\matrix {cos\theta_{S}cos\phi_{S} & -sin\phi_{S} & sin\theta_{S}cos\phi_{S}\cr
cos\theta_{S}sin\phi_{S} & cos\phi_{S} & sin\theta_{S}sin\phi_{S}\cr
-sin\theta_{S} & 0 & cos\theta_{S}}\right)
\left(\matrix {sin\theta_{0}^{'}cos\phi_{0}^{'}\cr
sin\theta_{0}^{'}sin\phi_{0}^{'}\cr
cos\theta_{0}^{'}}\right)
\end{equation}
As evident there will be many such values (one set for each time bin)
and their means and errors  may be determined.
\subsection{Determination of Spacecraft Attitude}
If $\theta_{v}$ is known, it is possible to determine the attitude of the
spacecraft as follows. This also requires two detectors-one aligned and
the other inclined (at a known angle) with the satellite's spin axis.

The equation connecting the unknown direction of the spin axis,
($\theta_{S}$ and $\phi_{S}$) to the known (true) direction cosines 
and known (apparent) direction cosines of
the source is the following.
\begin{equation}
\left(\matrix {sin\theta_{0}^{'}cos\phi_{0}^{'}\cr
sin\theta_{0}^{'}sin\phi_{0}^{'}\cr
cos\theta_{0}^{'}}\right) =
\left(\matrix {cos\theta_{S}cos\phi_{S} & cos\theta_{S}sin\phi_{S} & -sin\theta_{S}\cr
-sin\phi_{S} & cos\phi_{S} & 0\cr
sin\theta_{S}cos\phi_{S} & sin\theta_{S}sin\phi_{S} & cos\theta_{S}}\right)
\left(\matrix {sin\theta_{0}cos\phi_{0}\cr
sin\theta_{0}sin\phi_{0}\cr
cos\theta_{0}}\right)
\end{equation}
Three equations are obtained from the above matrix equation.
One of the equations is
\begin{equation}
-l_{1}sin\phi_{S}+l_{2}cos\phi_{S} = k_{2}
\end{equation}
where $l_{1}=sin\theta_{0}cos\phi_{0}$, $l_{2}=sin\theta_{0}sin\phi_{0}$,
and $k_{2}=sin\theta_{0}^{'}sin\phi_{0}^{'}$.
This leads to the quadratic equation
\begin{equation}
(l_{1}^{2}+l_{2}^{2})x^{2}+2k_{2}l_{1}x+(k_{2}^{2}-l_{2}^{2})=0
\end{equation}
Here $x=sin\phi_{S}$.
Solution of the last equation gives two values of $x$ from which the
value of $\phi_{S}$ may be obtained as $\phi_{S_1}=sin^{-1}(p_{1})$ and
$\phi_{S_2}=sin^{-1}(p_{2})$ where $p_{1}$ and $p_{2}$ are the two roots
of equation (21).

The other two equations are, respectively
\begin{equation}
l_{1}cos\theta_{S} cos\phi_{S}+l_{2}cos\theta_{S} sin\phi_{S}-l_{3}sin\theta_{S} = k_{1}
\end{equation}
and
\begin{equation}
l_{1}sin\theta_{S} cos\phi_{S}+l_{2}sin\theta_{S} sin\phi_{S}+l_{3}cos\theta_{S}= k_{3}
\end{equation}
Here $k_{1}=sin\theta_{0}^{'}cos\phi_{0}^{'}$, $l_{3}=cos\theta_{0}$
and $k_{3}=cos\theta_{0}^{'}$.
The last two equations give rise to the following equation
\begin{equation}
(k_{1}^{2}+k_{3}^{2})y^{2}+2k_{1}l_{3}y+(l_{3}^{2}-k_{3}^{2})=0
\end{equation}
Here $y=sin\theta_{S}$.
Solutions of the above equation gives two values of $y$ from which
the values of $\theta_{S}$ may be obtained as
$\theta_{S_1}=sin^{-1}(q_{1})$ and $\theta_{S_2}=sin^{-1}(q_{2})$
where $q_{1}=$ and $q_{2}=$ are the solutions of eqn. (24).

The correct values of $\theta_{S}$ and $\phi_{S}$ are to be decided
based on physical considerations.
\subsection{Three Axes Stabilized Spacecrafts}
In the case of a three axes stabilized spacecraft one may have a small
spinning platform on an extended boom on which the aligned and inclined
detectors may be placed. The above mentioned procedures may be utilised
to determine either source location or spacecraft attitude in this case
as well.
\section{Results:}
\subsection{The Timing Data and the Light Curve:}
Due to the non-availability of real data at hand, we take recourse to
generating data artificially.

The total duration of the time series data (detected counts vs. time) is equal
to 512 seconds with a time resolution of 256 ms (this is the width of each time
bin).

During the leading 51.2 seconds and the trailing 51.2 seconds the detected counts
are due only to the background. The background has a mean value of 1.8 counts
and is fluctuated according to a Poisson distribution.

Between the leading and trailing background data, in the remaining time interval
of 409.6 seconds a light curve is generated artificially. The light curve is
essentially triangular in shape. The total duration (409.6 seconds) of the
light curve is divided into two parts. The rising portion of the light curve has
a duration of 153.6 seconds. The decaying portion has a
duration of 256 seconds. Both the rising portion and the decaying portion of
the light curve are linear in shape. The rising portion has a slope of +3.8
while the decaying portion has a slope of -2.28. Poissonian fluctuations are
superimposed on the light curve generated by the Monte Carlo method (both in
the rising as well as in the decaying portion). This simulated light curve is
shown in Fig.2.
\begin{figure}[htp]
\includegraphics
[height=5cm,width=6.5cm,angle=0]{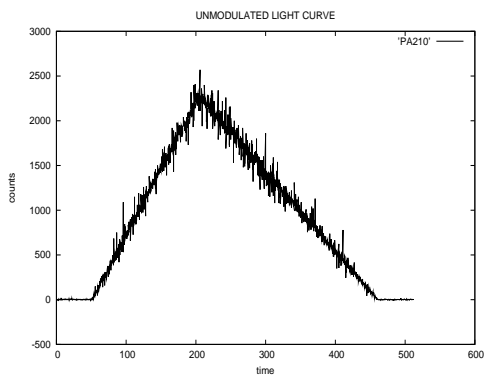}
\caption{The unmodulated light curve. The leading 51.2 seconds and the
trailing 51.2 seconds consist of only the background counts with an
average value of 1.8. Poissonian fluctuations are superimposed on the
background counts. The intervening 409.6 seconds constitute the simulated
light curve due to the high energy transient source. The rising portion
has a duration of 153.6 seconds and has a slope 0f 3.8. The decaying part
has a duration of 256 seconds and has a slope of -2.28.}
\end{figure}
\vskip 0.2cm
\subsection{The Modulated Light Curve:}
Due to the spinning motion of the satellite and the fact that the detector's
view axis is not aligned with the satellite's spin axis, the light curve of
the high energy transient (it is true also in the case of a steadily emitting
celestial object) will be modulated in amplitude. The modulation factor is
calculated as described earlier.

The modulated light curve corresponding to the original (unmodulated) light
curve (Fig.2) is shown in Fig.3. Here $\theta_{S}$ is the polar angle and
$\phi_{S}$ is the azimuth angle of the source as seen from
the spin-axis frame of reference. $\theta_{v}$ and $\phi_{v}$ are the
corresponding parameters for the view axis. To simplify the problem we
assume $\phi_{v} =0$. This essentially means a trivial rotation of the
spin axis coordinate system about its Z-axis by an amount equal to
$\phi_{v}$.
\begin{figure}[htp]
\includegraphics
[height=5cm,width=6.5cm,angle=0]{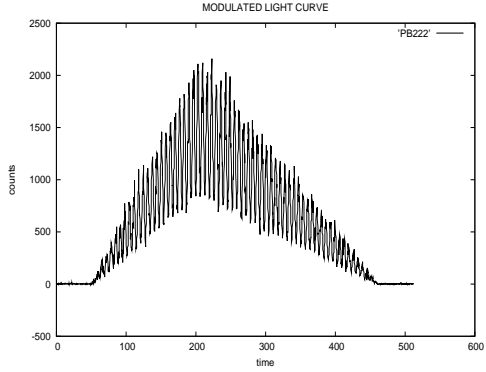}
\caption{The simulated light curve in Fig.1 of the high energy transient
modulated by the spin period of the satellite. The spin period is equal
to 11.78 seconds. The counts in the light curve are fluctuated according to
a Poisson distribution.}
\end{figure}
\vskip 0.2cm
The results of determination of $\theta_{v}$ are displayed in Table 1.
%{\bf Table 1}
%\begin{center}
%{\bf Table 1}
%\end{center}
%\centerline{\bf Table 1}
\begin{table}
\centerline{\bf Table 1}
\vskip 0.3cm
\caption{Estimated values of $\theta_{v}$ and their errors.}
\begin{center}
\begin {tabular}{llll}
Range of Data & True Value of & Estimated value & Error on  \\
(in cycles)   & $\theta_{v}$  &  of $\theta_{v}$ & $\theta_{v}$  \\
              & (in degrees)  & (in degrees)     & (in degrees) \\
  1-2              &  22.6         & 22.607     & 0.007            \\
&&&\\
  3-4              &  22.6         & 23.580     & 0.980             \\
&&&\\
  5-6              &  22.6         & 23.479     & 0.879             \\
&&&\\
  7-8              &  22.6         & 22.407     & -0.193               \\
&&&\\
  9-10             &  22.6         & 21.600     & -1.000              \\
&&&\\
 11-12             &  22.6         & 22.231     & -0.369               \\
&&&\\
\end{tabular}
\end{center}
\end{table}
The error on the mean of the estimated values of $\theta_{v}=0.283$
degrees which is equal to $1.25$ percent.
\section{Discussions:}
Obviously the amplitude of the modulation of a high energy transient
(or even a steady source) light curve depends on the parameters 
$\theta_{v}$, $\theta_{0}$ and $\phi_{v}$ etc (as shown in the figures 2 and
3).Since one assumes that the average of the term $cos\omega t$, $<cos\omega t>$
equals zero, one has to consider the time-history (light-curve) data only
for an integral number of the spin period of the satellite. In the present
work 2 cycles of data has been used for estimating each value of $\theta_{v}$.

It has been shown that even in the presence of fluctuations, this method
is still viable and yields quite good results.

However, the accuracy in the estimated value of $\theta_{v}$ depends on
the background. In another calculation where a background has been
assumed which is 13 times larger, the errors in the estimated value
of $\theta_{v}$ also becomes much larger, typically a few degrees.
Therefore, this method may be used effectively in only those cases where the
signal-to-noise ratios are quite large.

The spin period of the satellite and the integration time of the light curve
are very much realistic. Actually, the GRBM (Gamma Ray Burst Monitor)
on-board the SROSS C-2 satellite possessed an integration time of 256 ms
(Sinha, S et al) although the spin-period was nearly 12 seconds (about
half of the value used in the present work).

In the present work one assumes small (ideally point size) detectors since
the mathematical analysis does not assume any finite value of $\Delta\phi$
(due to the finite size of the detector). It is quite fortunate that
presently solid state (Si(PIN) and CdTe) detectors are available having size
as small as $1 mm X 1 mm$ that may be used to detect X-ray flares from the
Sun.
\section{Conclusions:}
The effects of modulation of light curves of celestial high energy photon
sources when the detector's view axis is not aligned with the spin axis of
the satellite has been described. It is shown how the orientation of the
view axis with respect to the spin axis may be determined by comparing this
modulated light curve with the light curve obtained from a detector whose
view axis is perfectly aligned with the spin axis (either in the same
satellite or in a different satellite). 
Finally, the usefulness of this apparently 
disadvantageous situation is described- how this effect may be utilised
in order to estimate the attitude of the spin axis or to determine the
unknown position coordinates of a celestial high energy photon source.
Further work in this direction will be reported shortly.
\section{Acknowledgements:}
I would like to express my deep sense of gratitude to all my revered
teachers.

I would like to thank the S.N. Bose Centre for financial support and
Ranjan Choudhury for mathematical help. Thanks are due to B S Acharya
for his very useful comments and suggestions to improve the
manuscript. I am indebted to my family
for their patience and support.

\end{document}